# Analysis of Differential Phase Shift Quantum Key Distribution

MONICA LAVALE

**Abstract:**
We review the implementation of two QKD protocols (BB84 and B92) keeping in mind that their implementations do not easily satisfy the requirement of use of single photons. We argue that current models do not take into account issues raised by the Uncertainty Principle related to time-location and transmission characteristics of single photons. This indicates that security proofs of current implementations even after the fixes for the recent successful hacks are made will be hard to obtain.

**Introduction**

Quantum computing provides procedures to solve computationally inefficient problems like prime factorization [3], database search [4], and cryptography, that is Quantum cryptography based key distribution (QKD) exploits the quantum properties of photons [1],[2],[5] and, in theory, provides a way of transmission of information in an unconditionally secure way over a network. There exist various types of QKD protocols (BB84[1], B92[2], Kak06 [5],[10]) that describe the optical apparatus to be used for encoding and decoding information in photons.

Classically, the one-time pad is the only unconditionally secure protocol of transmitting secret messages; however it has significant overhead when it comes to distributing the key to the receiver. A trusted courier is required to deliver the key to the receiver every time a message needs to be sent to the receiver. In addition, the key has to be of the same length as of the message and should not be reused to encrypt any other message in order to ensure the unconditional security of the algorithm. Quantum cryptography provides an entirely different solution to the cryptography problem by coding information in the form of states of photons that can be transmitted in an unconditionally secure way to the intended receiver. In quantum key distribution, quantum communication is used to exchange the key which is then used to encrypt the text message over a classical channel.

Oxborrow and Sinclair [6] argue that "no wholly consistent wave function for a single photon, whose amplitude squared gives the probability of the photon being located at a point in space (or momentum space), can be constructed." This indicates that implementation of effective single photon sources, which generate single photons at precise time instants, is going to be especially hard. Other problems are related to initialization of the system [11],[12]. It is for a combination of these reasons that implementations of BB84 have been hacked recently [8],[9].

In this paper, we first describe the B92 protocol (which is equivalent to the better known BB84 protocol), and the Photon Number Splitting attack that the protocol is most susceptible to, followed by a discussion of the challenges for the practical implementation for this protocol.

## Phase Modulation Protocol

The B92 protocol proposed by Bennett in 1992 [2]. It employs optical instruments like beam splitters, interferometers, photon detectors, phase modulator along with optical fibers in the apparatus for state preparation and transmission of photons. The sender of the key, Alice, encodes the bits of the key in two non-orthogonal states of the photon and sends it to the receiver, Bob. The certainty of secure, uninterrupted transmission relies on the Heisenberg's uncertainty principle which deduces that wave function of a single quantum system cannot be determined unless the preparation basis is known.

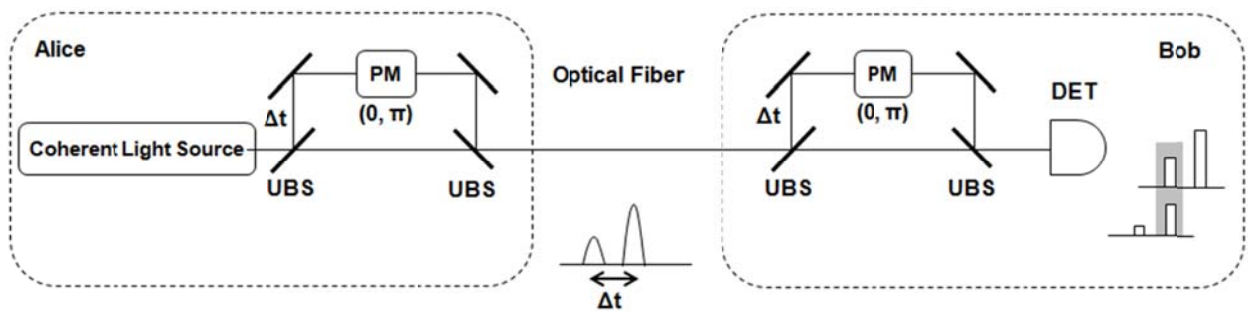

Figure 1: Optical apparatus for quantum key distribution as in B92 protocol. PM: Phase Modulator; UBS: Unbalanced Beam Splitter; DET: Detector

In her apparatus for state preparation, Alice uses a coherent source of light to produce an initial coherent pulse which is passed through an unbalanced interferometer. The beam splitter splits the pulse into two pulses such that the pulse which passes through the short arm of the interferometer is the bright reference pulse; and the pulse which passes through the long arm of the interferometer is the weak signal pulse. This weak signal pulse is phase shifted by either 0° or 180°, meaning information 0 and 1 respectively. The signal pulse and reference pulse are separated in time and the difference in their time is Δt. These encoded pulses are passed through an optical fiber to the other end where Bob receives the pulses and passes them through another interferometer. Since Bob has two pulses, each pulse is further split into two more pulses by the beam splitter. The detector connected to the interferometer detects pulses in three time slots. In the following sections, we cover all possibilities of phase shifting by Alice and measurement basis used by Bob. Since B92 uses only two non-orthogonal bases, there could be 4 possibilities:

1. Alice encodes bit 0 in photon state by using a phase shift of 0°; and Bob's measurement basis is correct, i.e., a phase shift of 0° in his apparatus. The schematic of the apparatus is shown in Figure 2.1 (a).

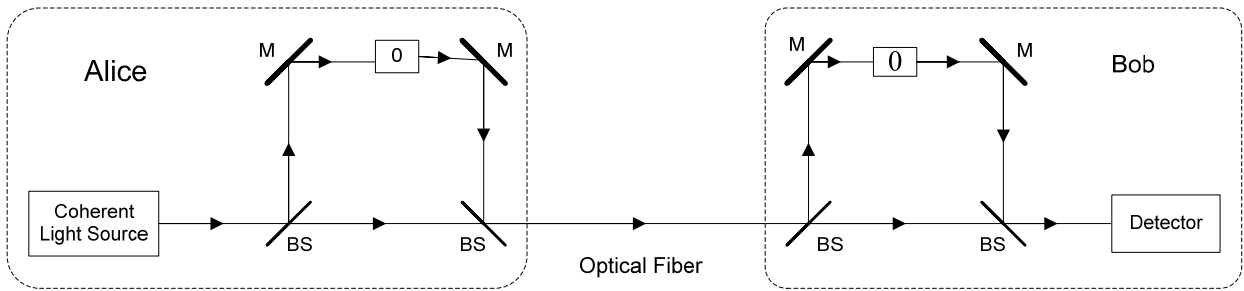

Figure 2.1 (a): Phase shift by Alice is 0° and phase shift by Bob is also 0°. BS: Beam Splitter; M: Mirror

Figure 2.1 (b) shows the waveforms of pulses occurring throughout the apparatus. The pulses in time slot $t_2$ cause constructive interference and results in a strong pulse shown in bold. The detector would measure the phase of this pulse and conclude that the information encoded in 0.

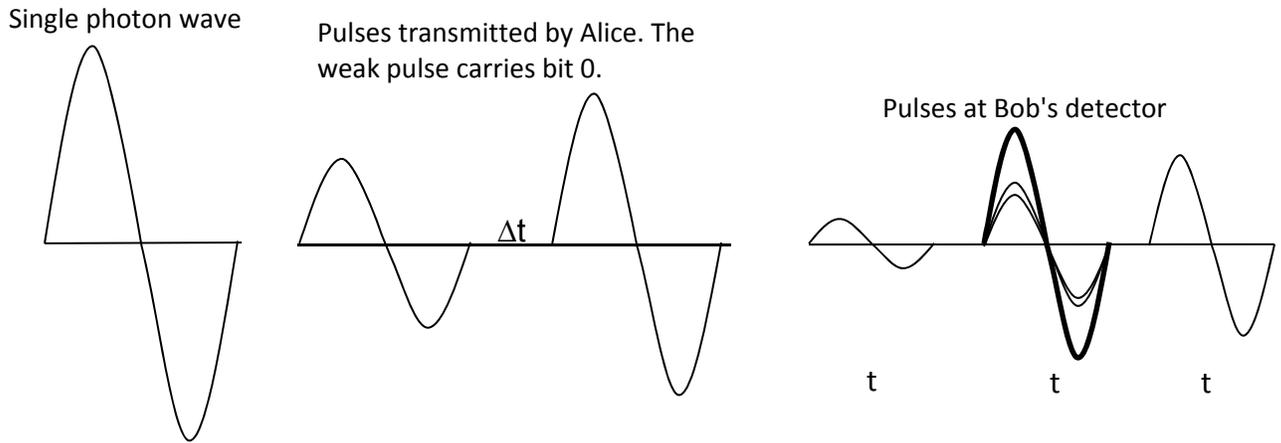

Figure 2.1 (b): Waveforms of pulses throughout the apparatus of Figure 2.1 (a)

2. The second case is that Alice encodes bit 0 in photon state by using a phase shift of 0˚; and Bob's measurement basis is incorrect, i.e., a phase shift of 180˚ in his apparatus. The schematic of the apparatus is shown in Figure 2.2 (a).

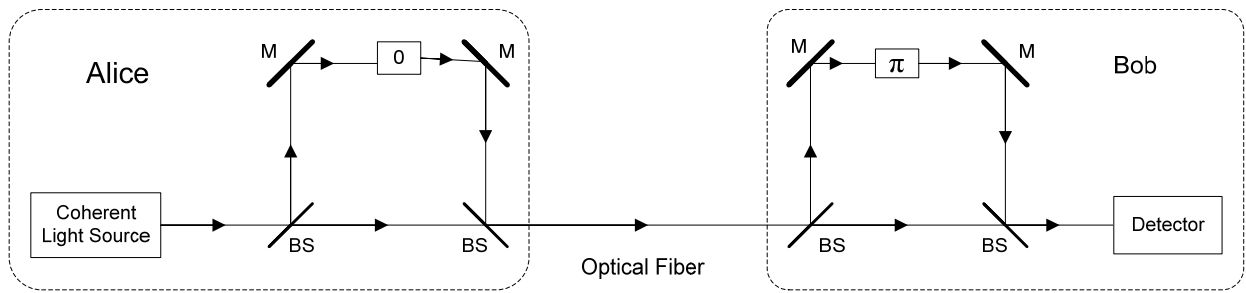

Figure 2.2 (a): Phase shift by Alice is 0˚ and phase shift by Bob is 180˚. BS: Beam Splitter; M: Mirror

Figure 2.2 (b) shows the waveforms of pulses occurring throughout the apparatus. The pulses in time slot $t_2$ cause destructive interference and hence no pulse is observed at the detector. This would signify that the measurement basis was wrong because of which information encoded is lost and we cannot conclude if that information was 0 or 1.

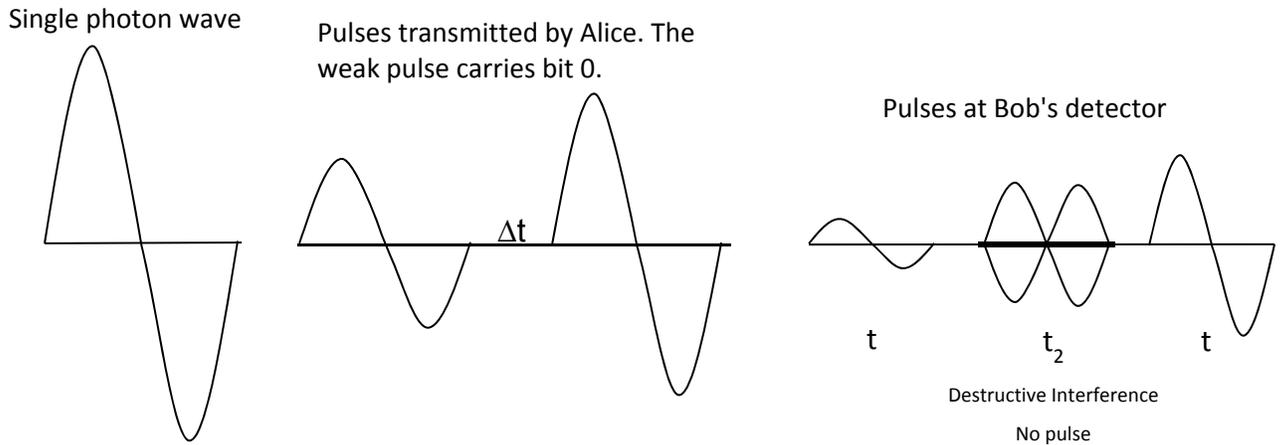

Figure 2.2 (b): Waveforms of pulses throughout the apparatus in Figure 2.2 (a)

3. The third case is that Alice encodes bit 1 in photon state by using a phase shift of 180˚; and Bob's measurement basis is correct, i.e., a phase shift of 180˚ in his apparatus. The schematic of the apparatus is shown in Figure 2.3 (a).

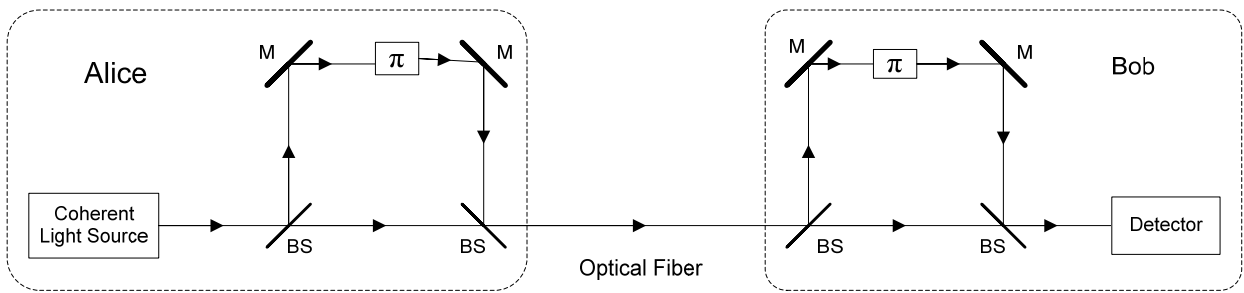

Figure 2.3 (a): Phase shift by Alice is 180˚ and phase shift by Bob is 180˚. BS: Beam Splitter; M: Mirror

Figure 2.3 (b) shows the waveforms of pulses occurring throughout the apparatus. The pulses in time slot $t_2$ are both phase shifted by 180˚ causing constructive interference and hence a strong pulse is observed at the detector. The detector measures the phase of the photon and we conclude that the information encoded was 1.

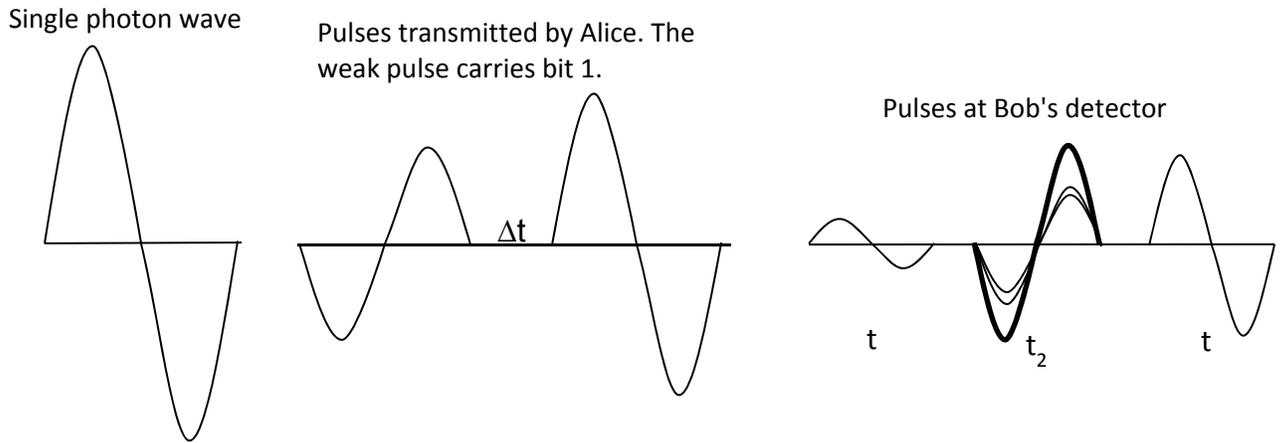

Figure 2.3 (b): Waveforms of pulses throughout the apparatus in Figure 2.3 (a)

4. The fourth case is that Alice encodes bit 1 in photon state by using a phase shift of 180°; and Bob's measurement basis is incorrect, i.e., a phase shift of 0° in his apparatus. The schematic of the apparatus is shown in Figure 2.4 (a).

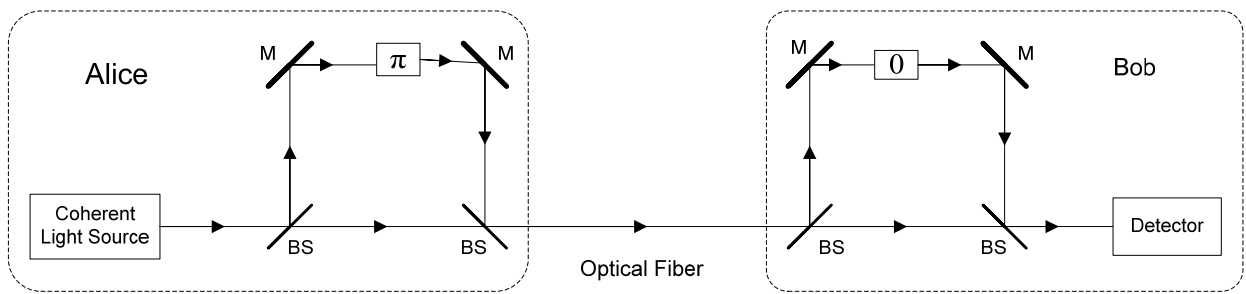

Figure 2.4 (a): Phase shift by Alice is 180° and phase shift by Bob is 0°. BS: Beam Splitter; M: Mirror

The waveforms of pulses occurring throughout the apparatus are same as shown in Figure 2.2 (b). The pulses in time slot $t_2$ cause destructive interference and hence no pulse is observed at the detector. This would signify that the measurement basis was wrong because of which information encoded is lost and we cannot conclude if that information was 0 or 1.

In this way, by observing the reading of detector in the time slot $t_2$, Bob informs Alice of the instances where he got correct results. Alice and Bob can then conclude to keep these bits as their key bits and discard all other bits. Table 1 shows the preparation basis and encoded values transmitted by Alice and the measurement basis used by Bob.

| Alice | Bit Value | 0 | 1 | 1 | 1 | 0 | 0 | 1 | 1 |
|---|---|---|---|---|---|---|---|---|---|
| | Phase | 0 | 180 | 180 | 180 | 0 | 0 | 180 | 180 |
| Bob | Basis | 0 | 0 | 180 | 0 | 180 | 180 | 180 | 0 |
| | Click | Y | N | Y | N | N | N | Y | N |
| | Bit Value | 0 | - | 1 | - | - | - | 1 | - |

Table 1: B92 protocol where Alice sends bits of 0 and 1 to Bob, Bob randomly measures the photons using base (0 for bit value 0 and 180 for bit value 1). The shared key bit is established at the Bob's detector as 011.

**PNS attack**

The basic assumption of the B92 protocol is that the coherent source of light emits one photon at a time and Alice encodes one bit of information in the states of photon by running it through a Phase Modulator. However, emission of a single photon from the light source is difficult to implement in practice [6]. As a consequence, multiple photons will be encoded with the same bits of information and sent over to Bob. If the eavesdropper gets access to the optical network, she can obtain the replicated information by siphoning some of the photons carrying same information to her optical apparatus. Some other photons would pass over to Bob without being disturbed. The schematic of this attack is shown in Figure 3.

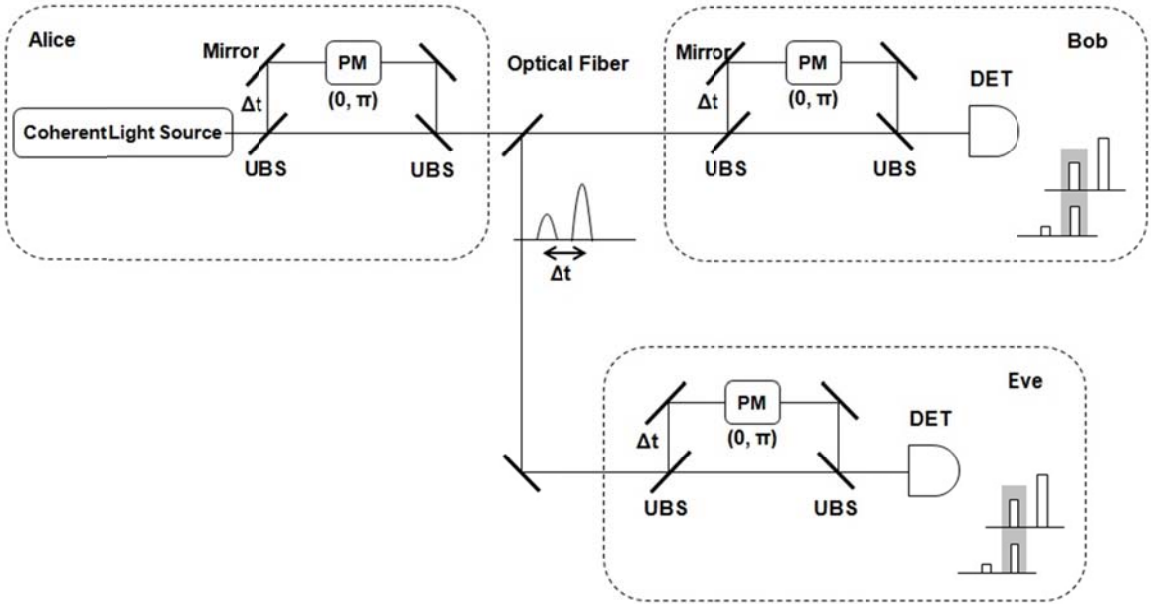

Figure 3: Photon Number Splitting attack by eavesdropper B92 protocol. PM: Phase Modulator; UBS: Unbalanced Beam Splitter; DET: Detector

**Implementation Challenges**

There are several challenges to the implementation of the apparatus for the encoding and transmitting photons. The first and important challenge is to design a single photon generator without which the system is prone to PNS attack. Although there is some progress in developing single photon generator that can be put in use for optical telecommunication [3],[7], the current engineering practice fall much short of perfect single photon sources [6].

The implementations by MagiQ and idQuantique use very low intensity pulses so that on an average there is a fraction of a single photon in one time slot. Nevertheless, the transmission will bunch several photons together in a Poisson distribution. Because of the fact that polarization states drift in optical fiber, the implementations use quantum phase modulation. When several photons are being transmitted instead of one, it is easier to find out the phase of the transmitted pulse. The industry implementations of BB84 and B92 protocols have recently been hacked [8],[9]. This suggests that the assumption that a very low power laser which implies a mean of a fraction of a single photon per unit of time does not provide sufficient security. The question of how this loss of security relates to the manner in which photons are detected with the assumption that their number follows the Poisson distribution needs to be worked out.

The other issue is that of noise. Even if single-photon sources existed, their use will be rendered ineffective by the fact that background noise requires that the signal strength be greater than it. Unless there is no background noise at all, one would be compelled to use several photons in the transmission which would then render the communication open to PNS attack. On the other hand, if a certifiably shielded, noise-free cable existed then there is no need for the use of quantum cryptography.

Another challenge is the transmission channel. If the photons were to be transmitted over fiber optics channel, the transmission length would be limited to about 100 km which is very less considering today's long distance communications. For lengths more than that, the photons would tend to change their phase because of constant reflections from inner coatings of fiber optics cable, and hence the information which was encoded in it would degrade. Since photon is the smallest quantum mechanical particle, it is very difficult to isolate it completely from the surrounding environment. This would obstruct its transmission over a wireless network since photons from the surrounding environment would interfere and cause some error in the information that the photon contains. Because of these reasons, practical implementations of the QKD protocols are limited to transmission over a few tens of kilometers; and still have the security hole for PNS attacks [4],[8],[9].

**Conclusions**

The apparatus of Figure 1, which is basic to the implementation of the B92 protocol, will also work for optical communication in the classical regime. This indicates that unless the signal source is certifiably quantum, the system can be considered to be classical and, therefore, subject to easy compromise. This leaves the Kak06 protocol as the most promising quantum cryptography protocol at this time [10] since it is more resistant to siphoning attacks compared to the other protocols. The robustness of the Kak06 protocol is due to the fact that Alice and Bob can use different polarization rotations on each qubit and, therefore, even if more than one photon is being transmitted for each qubit, Eve cannot use the information to break the code. The man-in-the-middle attack can be foiled by the use of trusted certificates [13].